\documentclass[11pt,a4paper]{article}

\usepackage{epsfig,amsmath,amssymb,cite}

\def\be{\begin{equation}}
\def\ee{\end{equation}}
\def\lp{\left(}
\def\rp{\right)}
\def\lb{\left[}
\def\rb{\right]}

\begin{document}

\title{Thin-shell wormholes in Brans--Dicke gravity} 
\author{Ernesto F. Eiroa$^{1,2,}$\thanks{e-mail: eiroa@iafe.uba.ar}\;, Mart\'{\i}n G. Richarte$^{1,}$\thanks{e-mail: martin@df.uba.ar}\; and Claudio Simeone$^{1,}$\thanks{e-mail: csimeone@df.uba.ar} \\
{\small $^1$ Departamento de F\'{\i}sica, Facultad de Ciencias Exactas y 
Naturales,} \\ 
{\small Universidad de Buenos Aires, Ciudad Universitaria Pab. I, 1428, 
Buenos Aires, Argentina}\\ 
{\small $^2$ Instituto de Astronom\'{\i}a y F\'{\i}sica del Espacio, C.C. 67, 
Suc. 28, 1428, Buenos Aires, Argentina}}

\maketitle

\begin{abstract}
Spherically symmetric thin-shell wormholes are constructed within the framework of Brans--Dicke gravity. It is shown that, for appropriate values of the  Brans--Dicke constant, these wormholes can be supported by matter satisfying the energy conditions. \\


\end{abstract}

\section{Introduction}

After the leading work by Morris and Thorne \cite{motho},  traversable Lorentzian wormholes  \cite{visser} have received considerable attention.  Such geometries  connect two regions of the same universe --or of two universes-- by a traversable  throat. A central objection against the existence of wormholes is that, within the framework of general relativity, the  flare-out condition \cite{hovis1} to be  satisfied at the throat  requires the presence of exotic matter, that is, matter which violates the energy conditions  \cite{motho,visser,hovis1,hovis2}. However, it was shown in Ref. \cite{viskardad}   that the amount of exotic matter necessary for the existence of a wormhole can be made infinitesimally small by  suitably choosing  the geometry, though it may be at the expense of large stresses at the throat \cite{dil,zas}. On the other hand, it was demonstrated that in some alternative theories of gravity the requirement of exotic matter can be avoided \cite{gra-wi,nts}; in particular, some years ago,  Anchordoqui {\it et al.} \cite{ancho} showed that, in Brans--Dicke gravity,  Lorentzian  wormholes of the Morris--Thorne type are compatible with matter which, apart from the Brans--Dicke scalar field, satisfies the energy conditions. Other related aspects of wormholes in Brans-Dicke or in scalar-tensor theories were also discussed in Refs. \cite{agnese,other}.

Thin-shell wormholes \cite{mvis} are mathematically constructed by cutting and pasting two manifolds to yield another one with a throat placed at the joining surface. The mechanical stability  and the matter content of spherically symmetric wormholes of this  kind have been studied by several authors, both within general relativity \cite{rg} and in alternative theories of gravity \cite{dil,alt,gra-wi}. Other geometries were also recently explored \cite{cil}. In the present work we apply the Darmois--Israel formalism  \cite{daris,mus} generalized to Brans--Dicke gravity \cite{dahia} for the construction of spherically symmetric thin-shell wormholes. We calculate the energy density and the pressure on the shell. We find that for certain values of the Brans--Dicke constant $\omega$, the wormhole radius can be chosen so that the matter on the shell satisfies the energy conditions, being the existence of the throat made possible by the presence of the  Brans--Dicke field. Throughout the paper units such that $c=G=1$ are used.

 \section{Thin-shells in Brans--Dicke gravity}

In the framework of present unified theories, a scalar field should exist besides the metric of the spacetime. Scalar-tensor theories of gravitation would be important when studying the early Universe, where it is supposed that the coupling of the matter to the scalar field could be nonnegligible.
In Brans--Dicke theory,  matter and non gravitational fields generate a long-range scalar field $\phi$ which, together with them, acts as a source of gravitational field. The metric equations generalizing those of general relativity are 
\be
R_{\mu\nu}-\frac{1}{2}g_{\mu\nu}R=\frac{8\pi}{\phi} T_{\mu\nu}+\frac{\omega}{\phi^2}\phi_{,\mu}\phi_{,\nu}-\frac{\omega} {2\phi^2} g_{\mu\nu}\phi_{,\alpha}\phi^{,\alpha}+\frac{1}{\phi}\phi_{;\mu;\nu}-\frac{1}{\phi}g_{\mu\nu} \phi_{:\alpha}^{;\alpha},
\ee 
where $R_{\mu\nu}$ is the Ricci tensor, $T_{\mu\nu}$ is the  energy-momentum tensor of  matter and fields --not including the  Brans--Dicke  field-- and $\omega$ is a dimensionless constant.
The field $\phi$ is a solution of the equation 
\be
\phi_{;\mu}^{;\mu}=\frac{1}{\sqrt{-g}}\frac{\partial}{\partial x^\mu}\left(\sqrt{-g}\ g^{\mu\nu}\frac{\partial\phi}{\partial x^\nu}\right)=\frac{8\pi T}{3+2\omega},
\ee
where $T$ is the trace of $T_\nu^{\mu}$. In the limit $\omega\to\infty$ the Einstein equations are recovered, provided that $\phi(\omega\to\infty)=1/G=1$. In Brans--Dicke theory, the junction conditions  across a smooth timelike hypersurface $\Sigma$ in the four dimensional manifold, can be obtained by projecting  on $\Sigma$ the equations above. The extrinsic curvature associated with the two sides of $\Sigma$  in terms of the unit normals $n^{\pm}_\gamma$ ($n^\gamma n_\gamma=1$) is given by
\begin{equation}
{K}^{\pm}_{ij} = -n^{\pm}_\gamma\left. \left(\frac{\partial^{2}X^\gamma}{\partial\xi^i\partial\xi^j}
                 +\Gamma^\gamma_{\mu\nu}\frac{\partial X^\mu}{\partial\xi^i}\frac{\partial X^\nu}{\partial\xi^j}\right)\right|_\Sigma,\end{equation}
where $X^{\gamma}$ are the coordinates of the four dimensional manifold, $\xi^i$ are the coordinates  on the hypersurface, and $\Gamma^\gamma_{\mu\nu}$ are the components of the connection associated with the metric $g_{\mu\nu}$. Then, with this definition, the  junction conditions in Brans--Dicke theory (generalized Darmois--Israel conditions) have the form \cite{dahia}
\begin{eqnarray}
    -\langle K^i_j\rangle+ \langle K\rangle\delta_j^i &=&  \frac{8\pi}{\phi}\lp S^i_j-\frac{S}{3+2\omega}\delta_j^i\rp,\label{cc}
   \\
   \langle \phi_{,N}\rangle  &=&   \frac{8\pi S}{3+2\omega},
\label{phin}
\end{eqnarray}
where the  notation $\langle \cdot \rangle$ stands for the jump of a given quantity across the hypersurface  $\Sigma$, $N$ labels the coordinate  normal to this surface and $S_j^i$ is the  energy-momentum tensor of matter and fields (except the field $\phi$) on the  shell located at $\Sigma$. The quantities $K$ and $S$ are the traces of $K^i_j$ and $S_j^i$ respectively.  The components of the metric and the Brans--Dicke field are continuous across the shell ($ \langle g_{\mu\nu}\rangle=0 $, $\langle \phi\rangle=0 $). Note that in the general relativity limit $\omega\to\infty$ the Lanczos equations     \cite{daris,mus} are recovered.

\section{Spherically symmetric wormholes}

Now let us apply the formalism introduced above to the construction of  spherically symmetric thin-shell wormholes. We start from the metric 
 \be
ds^2=- f(r)dt^2+ g(r)dr^2+ h(r)(d\theta^2+\sin^2\theta d\varphi^2),
\ee
so that $X=(t,r,\theta,\varphi)$. From this geometry we choose a radius $a$ and take two copies $\cal M^+$ and $\cal M^-$ of the region $r\geq a$, and paste them at the hypersurface $\Sigma$ defined by $r=a$, obtaining  a new manifold ${\cal M}={\cal M^+}\cup{\cal M^-}$. The radius $a$  is chosen so that there are no horizons and singularities in $\cal M$. If $h'(a)$ is positive  the flare-out condition is satisfied, and the resulting geometry describes a wormhole having a throat of radius $a$ connecting the two regions  $\cal M^+$ and $\cal M^-$. Introducing the coordinates 
$\xi=(\tau,\theta, \varphi)$ on $\Sigma$ (with $\tau$ the proper time on the shell),  the jump of the components of the extrinsic curvature is given by
\be
  \langle K_\tau^\tau\rangle = \frac{f'(a)}{f(a)\sqrt{g(a)}},
\ee
\be
\langle K_\theta^\theta\rangle  =   \langle K_\varphi^\varphi\rangle  =\frac{h'(a)}{h(a)\sqrt{g(a)}}.
\ee
Then the energy density $\sigma=-S_\tau^\tau$ and the pressures $p=S_\theta^\theta=S_\varphi^\varphi$ of the matter and fields in the shell (apart from the field $\phi$) are given by
\begin{eqnarray}
\sigma & = & -\frac{\phi(a)}{8\pi\sqrt{g(a)}}\lb\frac{2h'(a)}{h(a)}+\frac{1}{\omega}\lp\frac{f'(a)}{f(a)}+\frac{2h'(a)}{h(a)}\rp \rb , \label{sigma}\\
p & = & \frac{\phi(a)}{8\pi\sqrt{g(a)}}\lb \frac{f'(a)}{f(a)}+\frac{h'(a)}{h(a)}+\frac{1}{\omega}\lp\frac{f'(a)}{f(a)}+\frac{2h'(a)}{h(a)}\rp\rb ,\label{pe}
\end{eqnarray}
where $\phi(a)$ is the value of the Brans--Dicke field on the surface $\Sigma$. The constraint given by Eq. (\ref{phin}), which in this case takes the form $\langle \phi_{,N}\rangle=8\pi(2p-\sigma)/(3+2\omega)$, should be also satisfied. Consequently, we have 
\be
\sigma+p  =  \frac{\phi(a)}{8\pi\sqrt{g(a)}}\lb\frac{f'(a)}{f(a)}-\frac{h'(a)}{h(a)}\rb.\label{sigma+p}
\ee
Non exotic matter should satisfy the weak energy condition\footnote{The weak energy condition (WEC) states that for any timelike vector $T_{\mu\nu}V^\mu V^\nu\geq 0,$ which means that the local energy density as measured by any timelike observer is positive. In terms of the principal pressures it takes the form $\rho\geq 0$, $\rho+p_j\geq 0$ $\forall j$. The WEC implies the null energy condition (NEC) $T_{\mu\nu}k^\mu k^\nu\geq 0$ for any null vector, which in terms of the principal pressures takes the form $\rho+p_j\geq 0$ $\forall j$.}, i.e. $\sigma \geq 0$ and $\sigma+p\geq 0$; this condition would be violated if the flare-out condition is to be fulfilled within pure general relativity, which is easy to see from Eq. (\ref{sigma}) taking the limit $\omega\to\infty$.

\begin{figure}[t!]
\centering
\includegraphics[width=14cm]{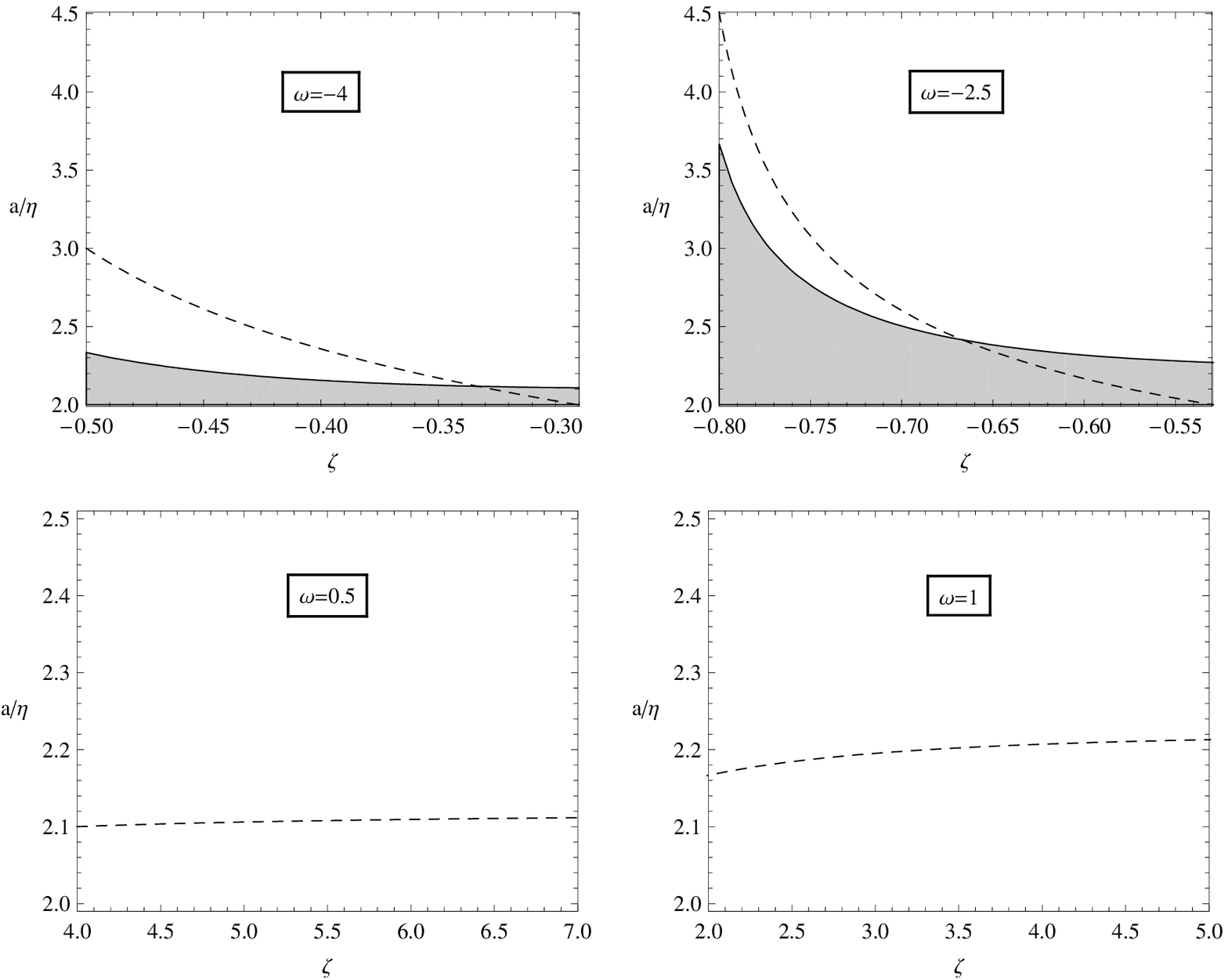}
\caption{The dashed line corresponds to values of the Brans--Dicke constant $\omega$ and parameters $\zeta$  and $\eta$ for which wormholes with throat radius $a$ exist. In the grey zone the condition $\sigma\geq 0$ is satisfied by the matter on the shell.}
\end{figure}

\begin{figure}[t!]
\centering
\includegraphics[width=14cm]{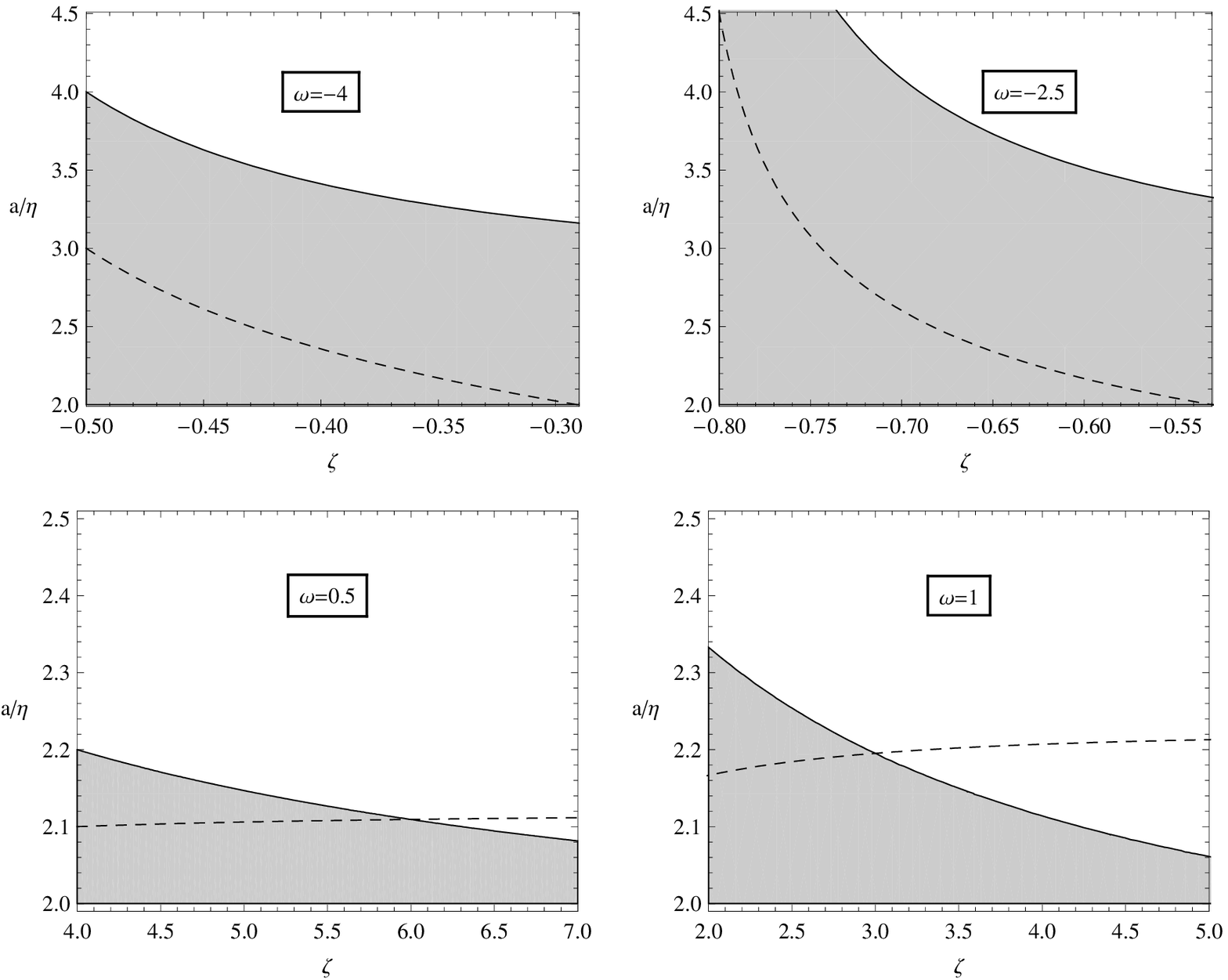}
\caption{The dashed line corresponds to values of the Brans--Dicke constant $\omega$  and parameters $\zeta$  and $\eta$ for which wormholes with throat radius $a$ exist. In the grey zone the condition $\sigma+p\geq 0$ is satisfied by the matter on the shell.}
\end{figure}

\begin{figure}[t!]
\vspace{1cm}
\centering
\includegraphics[width=14cm]{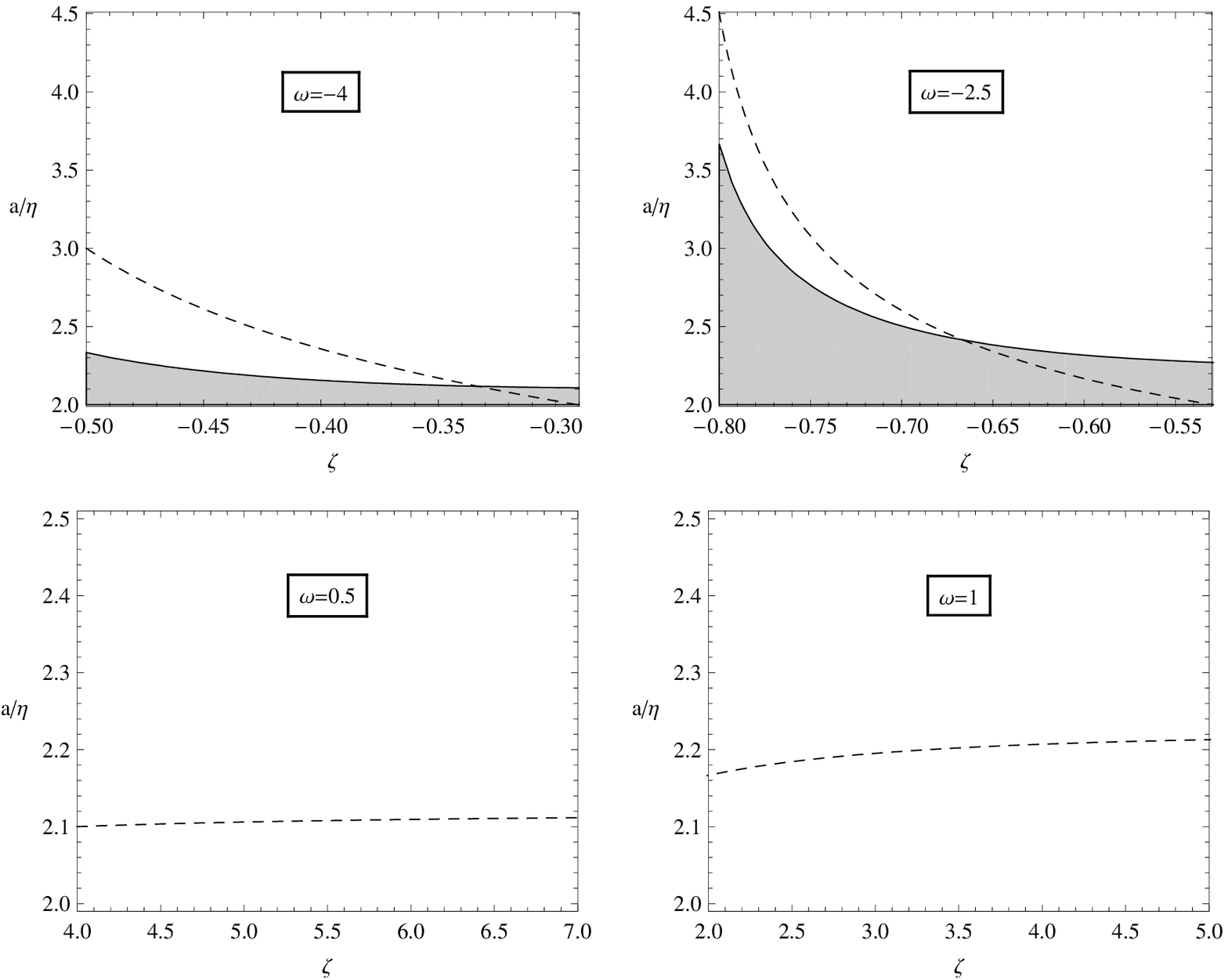}
\caption{The dashed line corresponds to values of the Brans--Dicke constant $\omega$  and parameters $\zeta$  and $\eta$ for which wormholes with throat radius $a$ exist. In the grey zone the weak energy condition ($\sigma\geq 0$ and $\sigma+p\geq 0$) is satisfied by the matter on the shell.}
\end{figure}

As a particular case, we consider  the spherically symmetric vacuum solution of Brans--Dicke equations (see \cite{agnese} and references therein), in which the  functions $f$, $g$ and $h$ are given by
\begin{eqnarray}
f(r)&=&\lp 1-\frac{2\eta}{r}\rp^A,\\ 
g(r)&=&\lp 1-\frac{2\eta}{r}\rp^B,\\  
h(r)&=&\lp 1-\frac{2\eta}{r}\rp^{1+B}r^2,
\end{eqnarray}
and the field $\phi$ takes the form
\be
\phi(r)=\phi _{0}\lp 1-\frac{2\eta}{r}\rp^{-(A+B)/2},
\ee
where
\be
\phi _{0}=\frac{4+2\omega}{3+2\omega},\ \ \ \ \ \
A=\frac{1}{\lambda },\ \ \ \ \ \ B=-\frac{\zeta+1}{\lambda},\ \ \ \ \ \
\lambda= \sqrt{(\zeta+1)^2-\zeta \lp 1-\frac{\omega \zeta }{2}\rp },
\nonumber
\ee
with $\eta >0$ and $\zeta$ constants. If the field $\phi $ is not constant, the geometry presents a naked singularity with radius $r_s=2\eta $. To construct the wormhole, we take $\omega <-2$ or $\omega >-3/2$ so that $\phi _{0}>0$, and a throat radius larger than  $r_s$. As $\lambda $ should be real and non zero, it follows that $(-1-\sqrt{-3-2\omega })/(2+\omega)<\zeta <(-1+\sqrt{-3-2\omega })/(2+\omega)$ if $\omega <-2$ and that $\zeta $ can take any value if $\omega >-3/2$. In the case $B+1\ge 0$ we have that $h'(a)>0$ $\forall a>2\eta $ and the flare out condition is satisfied; it happens when $2/\omega <\zeta <0$, for $\omega <-2$, and when $\zeta <0$ or $\zeta >2/\omega $, for $\omega >-3/2$. Replacing the explicit form of the metric and field,  we have
\begin{eqnarray}
\sigma  & = &  -\frac{\phi _{0}}{4\pi a^2}\lp \frac{a}{a-2\eta}\rp^{1+B+A/2}\lb 2\eta (B-1)+2a+ \frac{1}{\omega}\lp \eta A + 2\eta(B-1)+2a\rp\rb,
\label{sigma2}\\
p & = & \frac{\phi _{0}}{4\pi a^2}\lp \frac{a}{a-2\eta}\rp^{1+B+A/2}\lb\eta A+\eta (B-1)+a+\frac{1}{\omega}\lp \eta A + 2\eta(B-1)+2a\rp\rb,\label{p2}
\end{eqnarray}
with the constraint obtained from Eq. (\ref{phin}):
\be
2a+\eta \lb A+2B-2+\omega (A+B)\rb=0.
\label{constraint}
\ee
Then we obtain
\be
\sigma+p =  \frac{\phi _{0}}{4\pi a^2}\lp \frac{a}{a-2\eta}\rp^{1+B+A/2}\lb \eta A+ \eta (1-B)-a\rb.
\label{s+p}
\ee
The inequality $\sigma\geq 0$ is fulfilled when
\be
 \frac{a}{\omega}(\omega+1)+\frac{\eta}{\omega}\lb(B-1)(\omega+1)+\frac{A}{2}\rb\leq 0,
\label{sigma3}
\ee
 and the inequality  $\sigma+p\geq 0$ is satisfied if
\be
a \leq  \eta(A + 1-B);
\label{s+p2}
\ee
in both cases subject to the aditional condition given by Eq. (\ref{constraint}). If both inequalities above are satisfied the weak energy condition is not violated, and then the matter on the shell is not exotic. The throat radii such that conditions (\ref{sigma3}) and (\ref{s+p2}) are fulfilled are respectively displayed in figures 1 and 2 for some relevant values of the Brans--Dicke constant $\omega $ and the parameters $\zeta $ and $\eta $. The figure 3 shows where both conditions are simultaneously satisfied, so that configurations defined by such values contain non exotic matter in the shell placed at the wormhole throat. We see that this happens for $\omega <-2$, and  the radius $a$ slightly greater than $r_s=2\eta $. The range of the parameters for which exotic matter is not required grows as $\omega $ approaches $-2$.

\section{Discussion}

We have applied the generalized Darmois--Israel  formalism  to the construction of  spherically symmetric thin-shell wormholes in the framework of  Brans--Dicke gravity. We have obtained the energy density and the pressure of matter and fields other than $\phi$, in the shell placed at the  throat of the  wormhole. We have shown  that, for certain negative  values of the  Brans--Dicke constant, this matter satisfies the energy conditions if the throat radius is suitably chosen. Such values of the constant $\omega$ seem to be unphysical in the present Universe, but they could make sense in another scenario (i.e. far in the past). If the right hand side of Eq. (\ref{cc}) is understood as an effective source, the junction conditions present the same form as in general relativity:
 \be
-\langle K^i_j\rangle+ \langle K\rangle\delta_j^i  =  \frac{8\pi}{\phi}{\tilde S}^i_j,
\ee
with
\be
{\tilde S}^i_j =  S^i_j-\frac{S}{3+2\omega}\delta_j^i
\ee
the effective energy-momentum tensor on the surface.
With this definition, we would have that
\be 
\tilde \sigma= -{\tilde S}^\tau_\tau = -\frac{\phi(a)}{4\pi\sqrt{g(a)}}\frac{h'(a)}{h(a)}
\ee
is negative because of the flare-out condition. So, in this picture, the energy conditions are not satisfied by this effective surface tensor. The violation of these conditions comes from  the Brans--Dicke field, even in the presence of non exotic matter and other fields. This result is analogous to what was obtained by Anchordoqui {\it et al.} \cite{ancho} for wormholes of the Morris--Thorne type.

\section*{Acknowledgments}

This work has been supported by Universidad de Buenos Aires (UBA) and CONICET. Some calculations in this paper were done with the help of the package GRTensorII {\cite{grt}}. We want to thank Wolfgang Graf for pointing out errors in the previous version of the article.

\end{document}